\documentclass[prl,twocolumn,showpacs,floatfix,amsfonts]{revtex4}
\usepackage{graphicx}% Include figure files
\usepackage{bm}
\usepackage{amsmath}
\usepackage{amssymb}
\begin{document}
\preprint{}
\title{Field induced $d_{x^2-y^2}+id_{xy}$ state in $d$-density-wave metals}
\author{Jian-Xin Zhu}
\affiliation{Theoretical Division, MS B262,
Los Alamos National Laboratory, Los Alamos,
New Mexico 87545}
\author{A. V. Balatsky}
\affiliation{Theoretical Division, MS B262,
Los Alamos National Laboratory, Los Alamos,
New Mexico 87545}
%\date{\today}

\begin{abstract}
{ We argue that the $d_{xy}$ component of the order parameter  can be
generated to form the $d_{x^{2}-y^{2}}+id_{xy}$-density wave state by the
external magnetic field.  The driving force for this transition is the
coupling of the magnetic field with the orbital 
magnetism. The fully gapped
particle spectrum and the magnetically active collective mode of the condensate 
are discussed as a possible signature
of the $d+id^{\prime}$ density wave state. }
\end{abstract}
\pacs{74.20.-z, 74.25.Nf}
\maketitle

Recently, Chakravarty {\em et al.}~\cite{Chakra01}
proposed to model the pseudogap (PG)  of underdoped
 high-$T_{c}$ cuprates in the formation of a new
order ---$d$-density wave (DDW) state, which breaks the parity and time-reversal 
symmetry, and the
invariance of translation by one lattice constant and rotation by $\pi/2$. The 
scenario is
supported by the experiments showing that the PG and the superconducting gap 
coexist distinctly
below $T_c$. Although the debate about the mechanism for the PG is far from 
settled, it is
suggestive that the typical charactistics of the PG as observed in many 
experiments, including
photoemission~\cite{Norman98}, tunneling~\cite{Kras00}, muon spin 
relaxation~\cite{Sonier01}, can
be explained by the DDW model. More recently, the detection of the DDW ordering 
using impurity
resonance has been proposed~\cite{Zhu01c}.

We are going to argue here that,
 in addition to the dominant
$d_{x^2-y^2}$ ($d$) component of the DDW order parameter, a subdominant $d_{xy}$ 
($d^{\prime}$)
component can be generated by the magnetic field. We find that i) the presence 
of the additional
gap component $d^{\prime}$ will lead to the fully gapped quasiparticle spectrum; 
ii) the inherently
two component order parameter $d+id^{\prime}$ of the ddw phase in the field will 
exhibit
collective mode that can be excited in an out-of-plane ac magnetic field. Both 
of these features
can be used to distinguish between ddw scenario and alternatives. We discuss 
experimental
consequences below. The existence of the field induced $d+id^{\prime}$ 
superconducting state in
high-$Tc$ cuprates has recently been intensively studied.

The existence of the field induced $d+id^{\prime}$ superconducting state in 
high-$Tc$ cuprates has
recently been intensively studied. It was first proposed by 
Laughlin~\cite{Laughlin98} and
Ramakrishnan~\cite{Ramakrishnan98}, to explain the kink behavior observed in the 
thermal transport
experiment~\cite{Krishana97}, that the magnetic field can drive the original 
$d$-wave ordering
into the $d+id^{\prime}$ state. In the original proposal, the generation of the 
$d^{\prime}$
component was thought to be through the first-order bulk phase transition, but 
the important role
of the vortices for the anomaly was also pointed out by other 
authors~\cite{Aubin99,Ando00,Wang99}. Later
on it was argued~\cite{Balatsky00} that even when the magnetic field is near the 
upper critical
field, the $d$-wave state can be ``distorted'' by the external field, leading to 
a bulk
$d+id^{\prime}$-wave state with the intrinsic orbital moment. There is a 
significant distinction
between the $d$-wave superconducting (DSC) state and the DDW state: The magnetic 
induction
  in the superconducting state is inhomogeneous. In the DSC state, the 
electromagnetic
$U(1)$ gauge invariance is broken so in  an external magnetic field, the system 
is in  either the
Meissner state (type-I superconductors), where the magnetic flux is expelled 
from its bulk region,
or nucleating an array of vortices (type-II superconductors).
 The DDW state does 
not break the
gauge invariance and has  no analog of the Messner effect or the Abrikosov 
vortices. It is
then expected that  magnetic field inside a sample in the DDW state is 
homogeneous. In view of
this distinction, we argue that there should be an instability of the 
$d$-density wave state into
the $d+id^{\prime}$-density wave state even in the presence of a weak
magnetic magnetic field.

As in the DSC state, there also exists low-lying quasiparticle  states
around the 
nodes of the
$d$-density wave energy gap. At these nodes, the low-lying quasiparticles have 
vanishingly small
energy gap and they are most sensitive to an external electromagnetic 
perturbation, which opens
the possibility of the generation of the second component of the order 
parameter, orthogonal to
the initial $d$-density wave state. Second component of the
 order parameter  can 
be of $s$-wave or
$d^{\prime}$-wave orbital symmetry. Although, the $s$-wave component may be 
induced due to the
scattering at surfaces or interfaces, the $d^{\prime}$-wave order parameter is 
likely to be
generated  in the bulk sample when an external magnetic field is applied. The 
physical origin of
this instability is the bulk orbital magnetic moment $\langle M_{z}\rangle $ in 
the $d+id^{\prime}$ state. When an external magnetic field $H$ is applied 
perpendicular to the plane
of the two-dimensional (2D) system under consideration (namely, 
$\mathbf{H}\parallel
\hat{\mathbf{z}}$), the resulting coupling of the magnetic induction $B$ with 
the orbital magnetic
moment, $-\langle M_{z}\rangle B$, lowers the system free energy. As mentioned 
above, since in the
DDW state there is  no screening effect on the magnetic field, the magnetic 
induction $B$ is
homogeneous throughout the system and is close to the external magnetic field 
$H$. In the absence
of the magnetic field, the pure $d$-density wave state can be regarded as the 
equal admixture of
the orbital angular moment $L_{z}=\pm 2$ pairs:
\begin{equation}
W_{0}(\Theta)=iW_{0}\cos(2\Theta)=\frac{iW_{0}}{2}
[\exp(2i\Theta)+\exp(-2i\Theta)]\;.
\end{equation}
Here we have made an approximation to the order
parameter $W_{0}(\mathbf{k})\propto 
\langle c_{\mathbf{k}+\mathbf{Q},\sigma}^{\dagger}
c_{\mathbf{k},\sigma}\rangle\propto W_{0} (\cos k_{x}a -\cos k_{y}a)$ 
by confining the wave vector $\mathbf{k}$ near the Fermi surface and
introduced $\Theta$ as the 2D azimuthal
angle of the Fermi momentum, where $c_{\mathbf{k},\sigma}$ 
annihilates an electron of spin $\sigma$ at $\mathbf{k}$,  
$W_{0}$ is the magnitude of the pure $d$-wave component. In the presence
of an external magnetic field,
  the $L_{z}=\pm 2$ orbital wave functions becomes unequal
and the coefficients for them are shifted linearly with the magnetic field $H$:
\begin{eqnarray}
W_{0}(\Theta)&\rightarrow&  \frac{iW_{0}}{2}
[(1+\eta B)\exp(2i\Theta)+(1-\eta B)\exp(-2i\Theta)]
\nonumber \\
&=& i[W_{0}(\Theta)+iB W_{1}(\Theta)]\;,
\end{eqnarray}
where $B=H$ and $W_{1}\approx \eta \sin(2\Theta)$. Unlike the DSC,  the pure 
$d$-density wave order
parameter is imaginary while the field generated $d^{\prime}$-wave component is 
real. We will still
call the resulting DDW state as having the  $d+id^{\prime}$ symmetry because the 
relative phase
between the two components is still $\pi/2$ in the equilibrium.

To support the above physical intuitive, we provide a microscopic analysis in 
the following. The
$d$-density wave ordering comes from the formation of the particle-hole pairs 
caused by the
electron interactions.   The collective motion of these pairs can be represented 
by the
center-of-mass coordinates, while the relative motion by the relative 
coordinate. The structure of
the pair wavefunction is determined by the relative motion of two paired 
particles. In the
tight-binding model, both the DDW term and the coupling term between the 
magnetic field with the
electron orbital angular momentum (i.e., the orbital Zeeman coupling) 
are associated with the bilinear operator 
$c_{i,\sigma}^{\dagger}
c_{j,\sigma}$.
Therefore, in the presence of magnetic field,
both of these two terms should be assigned,  together with the kinetic energy 
term,  a phase
factor $\exp (\frac{i2\pi}{\Phi_{0}} \int_{\mathbf{R}_{j}}^{\mathbf{R}_{i}} 
\mathbf{A}\cdot
d\mathbf{l})$ to ensure the gauge invariance of the system Hamiltonian, where 
$\Phi_{0}=hc/e$ and
$\mathbf{A}$ is the vector potential.
This fact indicates
that the DDW order parameter is not influenced by the gauge phase factor, 
and allows us to treat the orbital Zeeman coupling effect
on the DDW order parameter independently.
Explicitly, the coupling of the magnetic field (parallel to $z$ direction) with 
the orbital
anugular momentum can be written as: $H_{B}=-i g\mu_{B}B 
\sum_{i,\bm{\delta},\sigma}
c_{i,\sigma}^{\dagger}\frac{1}{2a^{2}} (\mathbf{r}_{i}\times \bm{\delta})_{z}
c_{i+\delta,\sigma}$. Here $\bm{\delta}=(\pm a,0)$ or $(0,\pm a)$ is the
unit vector 
along the $x$ or
$y$ axis, $a$ is the lattice constant, $\mathbf{r}_{i}$ is the lattice vectors, 
$\mu_{B}$ is the
Bohr magneton, $g=2 m_e a^{2}t/\hbar^{2}$ with $t$ the nearest-neighbor hopping 
integral [the
energy unit used thereafter], $m_e$ the effective mass of electrons and $\hbar$ 
the Planck's
constant. In the momentum space, the system Hamiltonian can be written as:
\begin{eqnarray}
H&=&\sum_{\mathbf{k},\sigma} \xi_{\mathbf{k}} c_{\mathbf{k},\sigma}^{\dagger}
c_{\mathbf{k},\sigma} +\sum_{\mathbf{k},\sigma} [W_{\mathbf{k}}
c_{\mathbf{k},\sigma}^{\dagger}c_{\mathbf{k}+\mathbf{Q},\sigma}+\mbox{H.c.}]
\nonumber \\
&&-ig\mu_{B}B \sum_{\mathbf{k},\sigma} c_{\mathbf{k},\sigma}^{\dagger}
[\sin \mathbf{k}a \times \partial_{\mathbf{k}a}]_{z} c_{\mathbf{k},\sigma}
\;.
\label{EQ:Hamil}
\end{eqnarray}
Here $\xi_{\mathbf{k}}=-2[\cos k_x a+\cos k_y a]
-\mu$ with $\mu$ the chemical
potential is the single particle energy measured relative to the Fermi energy.
The DDW order parameter is given by $W_{\mathbf{k}}
=iW_{0}\varphi_{0}(\mathbf{k})
+W_1 \varphi_{1}(\mathbf{k})$, where
$\varphi_{0}(\mathbf{k})=\cos k_x a -\cos k_y a$ and
$\varphi_{1}(\mathbf{k})= \sin k_x a \sin k_y a$.   The amplitude of the
$d$- and $d^{\prime}$-wave compoponents $W_{0,1}$ are determined
self-consistently:
\begin{equation}
W_{0}=\frac{iV_{0}}{2N}\sum_{\mathbf{k}}
\langle c_{\mathbf{k}+\mathbf{Q},\sigma}^{\dagger}c_{\mathbf{k},\sigma}
\rangle \varphi_{0}(\mathbf{k})\;,
\label{EQ:W0}
\end{equation}
and
\begin{equation}
W_{1}=-\frac{2V_{1}}{N}\sum_{\mathbf{k}}
\langle c_{\mathbf{k}+\mathbf{Q},\sigma}^{\dagger}c_{\mathbf{k},\sigma}
\rangle \varphi_{1}(\mathbf{k})\;,
\label{EQ:W1}
\end{equation}
where $V_{0,1}$ are, respectively, the $d$- and $d^{\prime}$-channel 
interaction, $N$ is the
number of 2D lattice sites. That the $d$-wave component is imaginary is due to 
the equivalence of
$\mathbf{Q}=(\pi,\pi)$ and $-\mathbf{Q}$ enforced by the underlying band 
structure. The notation
$[\sin \mathbf{k}a \times \partial_{\mathbf{k}a}]_{z}$ represents $\sin (k_{x}a) 
\partial_{k_{y}a}
-\sin (k_{y}a) \partial_{k_{x}a}$.  We define
$\epsilon_{\mathbf{k}}=-2[\cos k_x 
a+\cos k_y a]$ so
that $\xi_{\mathbf{k}}=\epsilon_{\mathbf{k}}-\mu$. For $\mathbf{Q}=(\pi,\pi)$, 
we have following
symmetry properties: $\epsilon_{\mathbf{k}+\mathbf{Q}}=-\epsilon_{\mathbf{k}}$,
$\varphi_{0}(\mathbf{k}+\mathbf{Q})=-\varphi_{0}(\mathbf{k})$, and
$\varphi_{1}(\mathbf{k}+\mathbf{Q})=\varphi_{1}(\mathbf{k})$. In view of the 
fact that the DDW
state breaks the translational symmetry with lattice constant 
but conserves that 
by $\sqrt{2}a$
along the diagonals of the square lattice, it is convenient to halve the 
Brillouin zone, by
introducing  two kinds of electron operators $c_{\mathbf{k},\sigma}$ and
$c_{\mathbf{k}+\mathbf{Q},\sigma}$. 
The pairing of the particles and holes must cause correlations in their
relative motions. According to the structure of the Hamitonian and the
self-consistency conditions for the DDW order parameter, we can introduce
the following Green's functions to describe the correlation:
\begin{subequations}
\begin{eqnarray}
\mathcal{G}_{11}(\mathbf{k},\mathbf{k}^{\prime};\tau)&=&-\langle
T_{\tau} [c_{\mathbf{k},\sigma}(\tau)
c_{\mathbf{k}^{\prime},\sigma}^{\dagger}(0)]
\rangle \;,\\
\mathcal{G}_{12}(\mathbf{k},\mathbf{k}^{\prime};\tau)&=&-\langle
T_{\tau} [c_{\mathbf{k}+\mathbf{Q},\sigma}(\tau)
c_{\mathbf{k}^{\prime},\sigma}^{\dagger}(0)]
\rangle \;,\\
\mathcal{G}_{21}(\mathbf{k},\mathbf{k}^{\prime};\tau)&=&-\langle
T_{\tau} [c_{\mathbf{k},\sigma}(\tau)
c_{\mathbf{k}^{\prime}+\mathbf{Q},\sigma}^{\dagger}(0)]
\rangle \;,\\
\mathcal{G}_{22}(\mathbf{k},\mathbf{k}^{\prime};\tau)&=&-\langle
T_{\tau} [c_{\mathbf{k}+\mathbf{Q},\sigma}(\tau)
c_{\mathbf{k}^{\prime}+\mathbf{Q},\sigma}^{\dagger}(0)]
\rangle \;,
\end{eqnarray}
\end{subequations}
where the factor $T_{\tau}$ is a $\tau$-ordering operator as usual,
$c_{\mathbf{k},\sigma}(\tau)=e^{H\tau}c_{\mathbf{k},\sigma}e^{-H\tau}$ is the 
operator in the
Heisenberg representation. Given the Hamiltonian Eq.~(\ref{EQ:Hamil}), with aid 
of the equation of
motion for the field operator $c_{\mathbf{k},\sigma}(\tau)$ and
$c_{\mathbf{k},\sigma}^{\dagger}(\tau)$, and by performing a Fourier transform 
with respect to
$\tau$,
\begin{equation}
\mathcal{G}(\mathbf{k},\mathbf{k}^{\prime};\tau)
=k_{B}T \sum_{\omega_{n}}
\mathcal{G}(\mathbf{k},\mathbf{k}^{\prime};i\omega_{n})e^{-i\omega_{n}\tau}
\end{equation}
with $\omega_{n}=(2n+1)\pi k_{B}T$,
we establish a closed set of the
self-consistent equations for
$\mathcal{G}(\mathbf{k},\mathbf{k}^{\prime};i
\omega_{n})$, e.g.:
\begin{subequations}
\begin{eqnarray}
\delta_{\mathbf{k}\mathbf{k}^{\prime}}&=&\left(
i\omega_{n}-\xi_{\mathbf{k}+\mathbf{Q}}
+ig\mu_{B}B[\sin (\mathbf{k}+\mathbf{Q})a \times
\partial_{(\mathbf{k}+\mathbf{Q})a}]_{z}
\right)\nonumber \\
&& \times \mathcal{G}_{22}(\mathbf{k},\mathbf{k}^{\prime};i\omega_{n})
-2(W_{\mathbf{k}}^{*}+W_{\mathbf{k}+\mathbf{Q}})
\mathcal{G}_{21} (\mathbf{k},\mathbf{k}^{\prime};i\omega_{n})\;,\nonumber
\\
\end{eqnarray}
and
\begin{eqnarray}
0&=& \left(
i\omega_{n}-\xi_{\mathbf{k}}
+ig\mu_{B}B[\sin \mathbf{k}a \times
\partial_{\mathbf{k}a}]_{z}
\right) \mathcal{G}_{21}(\mathbf{k},\mathbf{k}^{\prime};i\omega_{n})
\nonumber \\
&& -2(W_{\mathbf{k}}+W_{\mathbf{k}+\mathbf{Q}}^{*}) \mathcal{G}_{22}
(\mathbf{k},\mathbf{k}^{\prime};i\omega_{n})\;.
\end{eqnarray}
\end{subequations}

To the approximation up to the first order in the
orbital-magnetic field coupling, we obtain
$\mathcal{G}_{21}(\mathbf{k},\mathbf{k}^{\prime};i\omega_{n})
=\mathcal{G}_{21}^{0}(\mathbf{k},\mathbf{k}^{\prime};i\omega_{n}) +\delta
\mathcal{G}_{21}(\mathbf{k},\mathbf{k}^{\prime};i\omega_{n})$, where
\begin{equation}
\mathcal{G}_{21}^{0}(\mathbf{k},\mathbf{k}^{\prime};i\omega_{n})
=\frac{(W_{\mathbf{k}}+W_{\mathbf{k}+\mathbf{Q}}^{*})
\delta_{\mathbf{k}\mathbf{k}^{\prime}} }
{D(\mathbf{k};i\omega_{n})}\;,
\end{equation}
and
\begin{eqnarray}
\delta\mathcal{G}_{21}(\mathbf{k},\mathbf{k}^{\prime};i\omega_{n})&=&
\frac{-ig\mu_{B}B (i\omega_{n}-\xi_{\mathbf{k}+\mathbf{Q}})}
{D(\mathbf{k};i\omega_{n})}
\nonumber \\
&& \times
[\sin \mathbf{k}a \times \partial_{\mathbf{k}a}]
\mathcal{G}_{21}^{0}(\mathbf{k},\mathbf{k}^{\prime};i\omega_{n})
\;.
\label{EQ:DELTA-G21}
\end{eqnarray}
where $D(\mathbf{k};i\omega_{n}) 
=(i\omega_{n}-E_{\mathbf{k},1})(i\omega_{n}-E_{\mathbf{k},2})$
with $E_{\mathbf{k},1(2)}=\pm\sqrt{\epsilon_{\mathbf{k}}^{2}+ \vert 
W_{\mathbf{k}} +
W_{\mathbf{k}+\mathbf{Q}}^{*} \vert^{2} }-\mu$. We take the ansatz  that $V_{0}$ 
is bigger than
$V_{1}$~\cite{Note1} such that in the absence of the magnetic field, the 
$d$-wave ordering is pure
and no secondary phase transition for the appearance of the $d^{\prime}$ 
ordering occurs.
Therefore, the DDW gap appearing in the $\mathcal{G}^{0}$ is,  $W_{\mathbf{k}} 
=iW_{0}
\varphi_{0}(\mathbf{k})$. The momentum dependence of $\varphi_{0}(\mathbf{k})$ 
leads to $[\sin
\mathbf{k}a \times \partial_{\mathbf{k}a}]
 \varphi_{0}(\mathbf{k})
=2\varphi_{1}(\mathbf{k})$.   %~\cite{Note2}.
Substitution of Eq.~(\ref{EQ:DELTA-G21}) into
Eq.~(\ref{EQ:W1}) yields:
\begin{eqnarray}
W_{1}&=&-\frac{4V_{1}}{N} \sum_{\mathbf{k}\in \mbox{rbz} }
\mbox{Re}[\delta\mathcal{G}_{21}(\mathbf{k},\mathbf{k};\tau=0)]
\varphi_{2}(\mathbf{k})
\nonumber \\
&=& \eta B W_{0}\;,
\label{EQ:INDUCED}
\end{eqnarray}
where
\begin{eqnarray}
\eta&=&-\frac{16g\mu_{B}V_{1}k_{B}T}{N} \sum_{\mathbf{k}\in \mbox{rbz} }
\sum_{\omega_{n}}
\frac{\epsilon_{\mathbf{k}}\varphi_{1}^{2}(\mathbf{k}) }
{D^{2}(\mathbf{k};i\omega_{n})}\nonumber \\
&\approx& \frac{16g\mu_{B}N(0)V_{1}}{E_{F}}
\;,
\end{eqnarray}
where $N(0)$ is the density of states at the Fermi energy $E_{F}$. By taking 
the Fermi wave length of a few 
lattice constant $a$ ($\sim 4$\AA) and
$N(0)\vert V_{1}\vert \sim 0.3$, 
it is estimated $\vert W_{1}/W_{0}\vert\approx 10^{-2}$ at $B=10 \mbox{T}$, 
which 
makes the amplitude of the induced component $\vert W_{1}\vert$ to be on the 
order of a few Kelvin. 
Eq.~(\ref{EQ:INDUCED}) shows microscopically that the magnetic field can drive
the pure $d$-wave state into the $d+id^{\prime}$
state. This analysis also indicates that the $d+is$-density wave state
could not be generated by the applied magnetic field, which is significantly
different from the situations at the sample surfaces or interfaces.

Equation~(\ref{EQ:INDUCED}) suggests that the phenomenological
Ginzburg-Landau (GL) free energy functional must contain the linear coupling
between the original $d$-density wave order parameter and the
field-induced $d^{\prime}$-density wave order parameter,
$f_{\mbox{int}}=i\frac{\eta}{2}(iW_{0}) W_{1}^{*} B +\mbox{c.c.}$.
Consequenetly, we can write down the system GL functional of the form:
\begin{eqnarray}
\mathcal{F}&=&\int d^{2}r [ \frac{\alpha_{0}}{2}(T-T_{c}^{0})
\vert W_{0}(\mathbf{r})\vert^{2}
+\frac{\beta_{0}}{4}\vert W_{0}(\mathbf{r})\vert^{4}
\nonumber \\
&&+\frac{K_{0}}{2}\vert \nabla (iW_{0}(\mathbf{r}))\vert^{2}
+\frac{K_{1}}{2}\vert \nabla W_{1}(\mathbf{r})\vert^{2}
+\frac{\alpha_{1}}{2}\vert W_{1}(\mathbf{r})\vert^{2}
\nonumber \\
&& +f_{\mbox{int}}(\mathbf{r})
]\;,
\label{EQ:GL}
\end{eqnarray}
where the first two terms describe the instability of the pure $d$-density wave 
state, with
$T_{c}^{0}$ being the transition temperature in the absence of magnetic field. 
The  last two terms
represent the energy shift of the $d$-wave state as a result of the 
field-induced
$d^{\prime}$-wave order parameter, where $\alpha_{1}$ is positive. Notice that, 
unlike the
superconducting order parameter, the gradient operator on the DDW order 
parameter is not shifted
by the vector potential
 because the DDW pairs do not carry charge.
It follows from Eq.~(\ref{EQ:GL}) that, as far as the $d^{\prime}$-wave 
component is concerned,
the coupling to the magnetic field term [i.e., $f_{\mbox{int}}$] is linear, 
while the stiffness
term [i.e., the second last term in Eq.~(\ref{EQ:GL})] is quadratic. Therefore, 
at least at the
weak field so that $W_{1}$ is small, the linear term is dominant. Therefore, the 
system gains
energy by having a nonzero equilibrium value of $W_{1}$. By treating $W_{1}$ and 
$W_{1}^{*}$ as
independent variables, the GL functional $\mathcal{F}$ is minimized by enforcing 
$\frac{\delta
\mathcal{F}}{\delta W_{1}} =\frac{\delta \mathcal{F}}{\delta W_{1}^{*}}=0$, 
which leads to
\begin{equation}
W_{1}=\frac{\eta B}{\alpha_{1}}W_{0}\;.
\end{equation}
Upon  substituting  the above result into Eq.~(\ref{EQ:GL}), we find  the energy 
gained by the
system with the induced $d^{\prime}$-density wave component: $\delta 
\mathcal{F}=-\int d^{2}r
\eta^{2} \vert W_{0}\vert^{2} B^{2}/2\alpha_{1} $. Therefore, the transition 
temperature which is
now field dependent, and is renormalized by the magnetic field as: 
$T_{c}(B)=T_{c}^{0}+\delta
T_{c}(B)$, where $\delta T_{c}(B)=\eta^{2}B^{2}/2\alpha_{0}\alpha_{1}$. It then 
follows that
coupling of the magnetic field with the orbital angular momentum shifts the 
instability of the
$d$-density wave ordering to the high temperature as schematically shown in the 
$(T,B)$ phase
diagram. Here we note that, since the particle-hole pairing takes place
with the equal spin, the coupling
between the magnetic field and the electron spin (i.e., the spin Zeeman
coupling) will not depress the induction of $d^{\prime}$ component
in the DDW metal. We do not address here the effect of very strong field
that could ultimately supress the DDW state.

\begin{figure}
\includegraphics{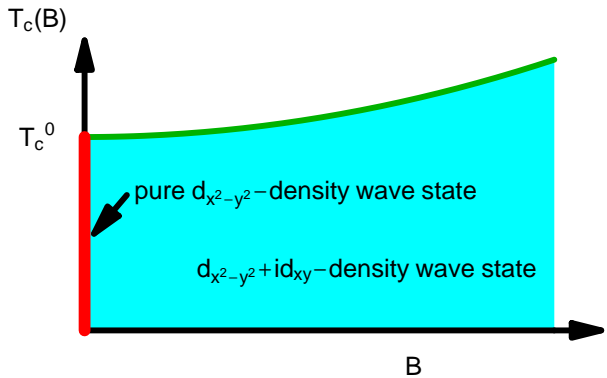}
\caption[*]{The schematic drawing of the $(B,T)$ phase diagram of the 
$d$-density wave metal. The
pure $d_{x^{2}-y^{2}}$ density wave state is marked by the thick solid line 
lying on the $y$ axis.
Quite different from the case of $d$-wave superconductors, 
the $d_{x^{2}-y^{2}}$-density wave state has
the instability into the $d_{x^{2}-y^{2}}+id_{xy}$-wave state 
in the presence of an arbitrary  magnetic field coupled 
to the electron orbital angular momentum. 
} 
\label{FIG:Phase-Diagram}
\end{figure}

Up to now our analysis of the induction of the secondary $d^{\prime}$ component
has been focused on the equilibrium solution. If we assume that this secondary
$d^{\prime}$ order parameter has been created, we can write in general
$iW_{0}=\vert W_{0}\vert e^{i\phi_0}$ and
$W_{1}=\vert W_{1} \vert e^{i\phi_{1}}$, and study the dynamics of the
relative phase $\phi=\phi_{1}-\phi_{0}$, which is governed
by~\cite{Balatsky00b}:
\begin{equation}
\frac{\partial^{2} \phi}{\partial t^{2}}=-\rho^{-1}
\frac{\delta \mathcal{F}}{\delta \phi}\;,
\label{EQ:PHASE}
\end{equation}
where $\rho^{-1}\approx N(0)$. With
Eq.~(\ref{EQ:GL}), we find
\begin{equation}
\frac{\partial^{2} \phi}{\partial t^{2}}=
-\rho^{-1}\eta B\vert W_{0}\vert \vert W_{1} \vert \cos\phi
-s^{2} \nabla^{2}\phi\;,
\end{equation}
which leads to the clapping mode with dispersion 
$\omega^{2}(B,k)=\omega_{0}^{2}(B)+s^{2}k^{2}$
with $\omega_{0}^{2}(B)=\eta B^{2} \vert W_{0}\vert^{2}/\rho$ and $s^{2}=\vert 
W_{0}\vert^{2}(K_{0}
+\eta^{2}B^{2}K_{1})/4\rho$. This mode represents the oscillation of the 
relative phase between
the $d$ and $d^{\prime}$ components of the DDW order parameter, and is tunable 
by the magnetic
field.

\begin{figure}
\includegraphics{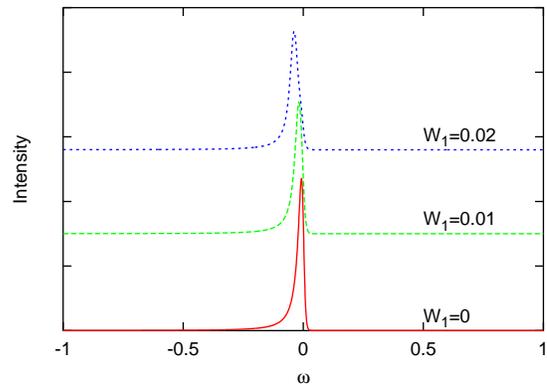}
\caption[*]{Energy distribution curves at $(\frac{\pi}{2},\frac{\pi}{2})$
for various values of the $d_{xy}$-wave component $W_{1}=0$ (solid line),
0.01 (dashed line), and 0.02 (dotted line). The other parameter values:
$\mu=0$ and $T=0.005$.
}
\label{FIG:ARPES}
\end{figure}

We thus have proved that (a) the applied magnetic field can generate the 
$d_{xy}$ order parameter
in the $d$-density wave metal, whose amplitude is linearly proportional to the 
field strength, (b)
the transition into the $d+id^{\prime}$-density wave state occurs at a higher 
transition
temperature, and (c) there exists a new clapping mode corresponding to the 
oscillation of the
relative phase between the two components. We now turn to the experimental 
observation of the
existence of the induced $d^{\prime}$ component.  Angle resolved photoemission 
spectroscopy (ARPES)
can be used to directly detect the existence of the induced $d^{\prime}$ 
component by measuring
the low-lying excitations at the nodal directions, where, near the half filling, 
the dominant
$d$-density wave gap closed at the Fermi surface while $d^{\prime}$-density wave 
gap reaches the
maximum. Figure~\ref{FIG:ARPES} displays the low-temperatures ARPES signal
$I(\mathbf{k},\omega)=f(\omega)A(\mathbf{k},\omega)$ at 
$\mathbf{k}=(\frac{\pi}{2},\frac{\pi}{2})$
for various values of the induced $d^{\prime}$-wave component $W_{1}$. Here  is 
the Fermi
distribution function $f(\omega)=1/[\exp(\omega/k_{B}T)+1]$ and the spectral 
function
$A(\mathbf{k},\omega)=2[\delta(\omega-E_{\mathbf{k},1}) 
+\delta(\omega+E_{\mathbf{k},2})]$. As is
shown, the spectral peak is shifted with the magnitude of $W_{1}$, which is in 
turn linearly
proportional to the magnetic field. As another consequence, in the presence 
of $W_{1}$, the quasiparticle spectrum is fully gapped. Therefore, 
if the magnetic field is applied perpendicularly to the 2D DDW metal,
the electronic specific heat at low temperatures would be exponentially
decaying.
%Nevertheless,
%although the magnetic field does not affect the
%DDW order parameter except generating the $d^{\prime}$ component,
%it directly makes the quasiparticles of the DDW state form
%Landau levels, which will
%complicate the interpretation of the ARPES data.

To conclude, we show that the $d$-density wave state has an
instability into the  $d+id^{\prime}$ ordering in the presence of
the magnetic field. The field induced $d^{\prime}$-wave
 component is proportional to the
field strength. The mechanism for the induction of
the $d^{\prime}$ component is purely the coupling between
the magnetic field and the orbital anugular momentum.

{\bf Acknowledgments}:  We thank J.C. Davis, M.J. Graf, R.B. Laughlin,
K.K. Loh, and D. Morr for stimulating discussions.
This work was supported by the Department of Energy through the Los Alamos
National Laboratory.

\end{document}